\documentclass[preprints,article,accept,moreauthors,pdftex]{Definitions/mdpi} 

\firstpage{1} 
\pubvolume{xx}
\issuenum{xx}
\articlenumber{xx}
\pubyear{2020}
\copyrightyear{2020}
\history{Received: 4 November 2020; Accepted: 24 November 2020; Published: date}

\pdfoutput=1

\usepackage{eurosym}
\usepackage{subcaption}
\usepackage{tabularx}

\Title{Optical Readout of the ARIADNE LArTPC using a Timepix3-based Camera}

\Author{Adam Lowe, Krishanu Majumdar, Konstantinos Mavrokoridis *, Barney Philippou, Adam Roberts *, Christos Touramanis and Jared Vann}

\AuthorNames{Adam Roberts, Jared Vann and Konstantinos Mavrokoridis}

\address{University of Liverpool, Department of Physics, Oliver Lodge Bld, Oxford Street, Liverpool, L69 7ZE, UK}

\corres{Correspondence: k.mavrokoridis@liverpool.ac.uk (K.Mv.); aroberts@hep.ph.liv.ac.uk (A.R.) }

\abstract{The ARIADNE Experiment, utilising a 1-ton dual-phase Liquid Argon Time Projection Chamber (LArTPC), aims to develop and mature optical readout technology for large scale LAr detectors. This paper describes the characterisation, using cosmic muons, of a Timepix3-based camera mounted on the ARIADNE detector. The raw data from the camera are natively 3D and zero suppressed, allowing for straightforward event reconstruction, and a gallery of reconstructed LAr interaction events is presented. Taking advantage of the 1.6~ns time resolution of the readout, the drift velocity of the ionised electrons in LAr was determined to be 1.608~$\pm$~0.005~mm/$\mu$s at 0.54~kV/cm. Energy calibration and resolution were determined using through-going muons. The energy resolution was found to be approximately 11~\% for the presented dataset. A preliminary study of the energy deposition ($\frac{dE}{dX}$) as a function of distance has also been performed for two stopping muon events, and comparison to GEANT4 simulation shows good agreement. The results presented demonstrate the capabilities of this technology, and its application is discussed in the context of the future kiloton-scale dual-phase LAr detectors that will be used in the DUNE programme.}


\begin{document}

\begin{center}
\includegraphics[width=0.35\textwidth]{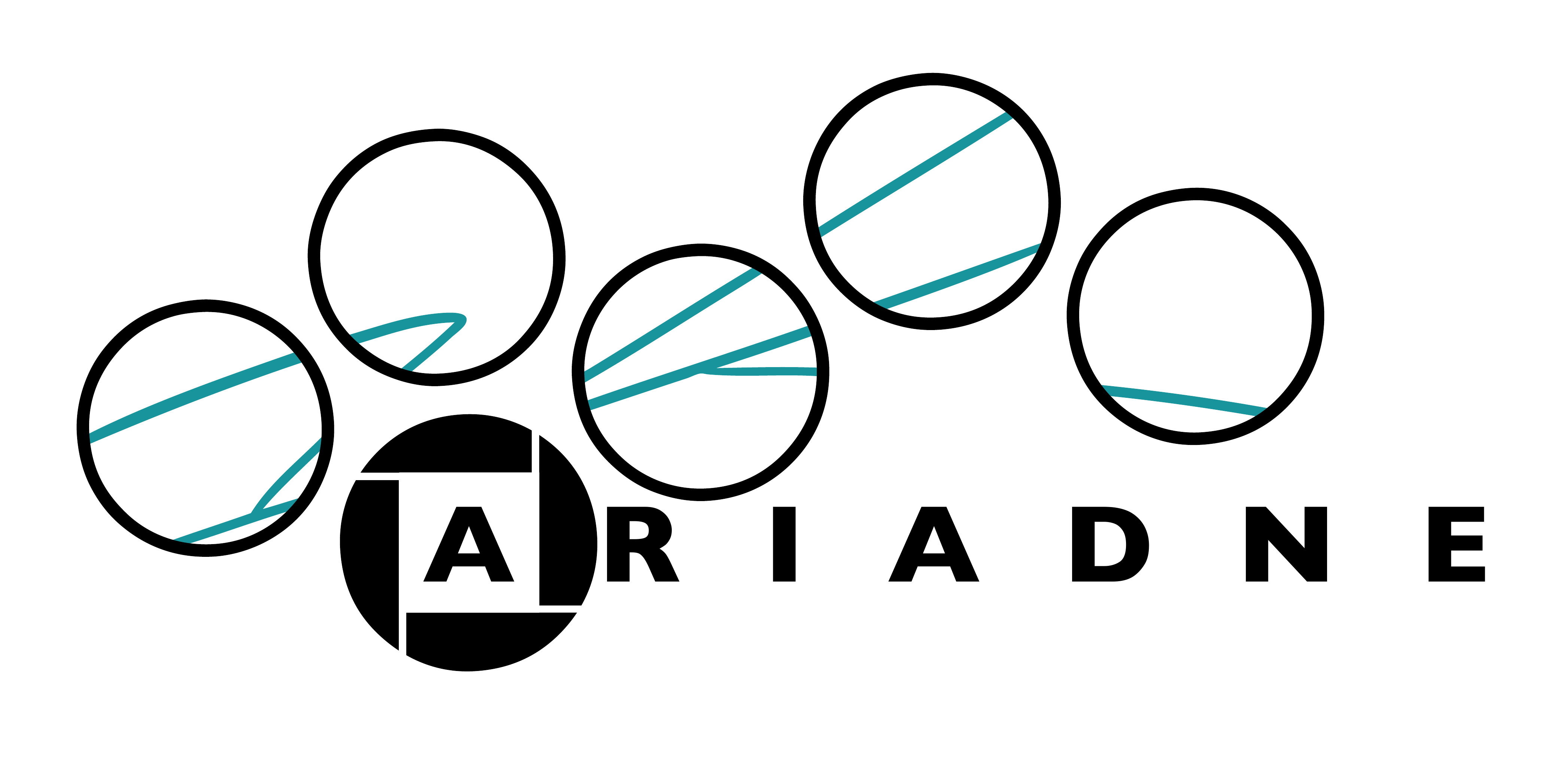}
\end{center}

\section{Introduction}
Liquid Argon Time Projection Chambers (LArTPCs) have been an indispensable type of particle detector for over 40 years, and have only continued to grow in size and sophistication. The current generation of neutrino detectors - such as SBND (112~tons), MicroBooNE (89~tons) and ICARUS-T600 (470~tons), which together make up the Short Baseline Neutrino Program~\cite{SBNP}, and the single- (411~tons) and dual-phase (300~tons) ProtoDUNE experiments \cite{ProtoDUNE-SP, ProtoDUNE-DP} - are already approaching the kiloton-scale. It is evident then that future LArTPCs in the neutrino sector will be able to reach the kiloton-scale - for example, four 17,000 ton LArTPCs have been proposed for use on the DUNE project \cite{DuneCDRVol1, DuneCDRVol2, DuneCDRVol3, DuneCDRVol4, DuneIDRVol1, DuneIDRVol2, DuneIDRVol3}.

Given the high construction and operating costs, as well as the sheer complexity, of such large detectors, early and innovative R\&D therefore has the potential for a large return on investment over an experiment's lifetime.
\\
\\
The ARIADNE (\textbf{AR}gon \textbf{I}m\textbf{A}ging \textbf{D}etectio\textbf{N} chamb\textbf{E}r) Experiment is based around a 1-ton dual-phase LArTPC. Using this detector, the project aims to demonstrate the feasibility of optical readout of LArTPCs on a large scale, and develop an ongoing program for the characterisation and maturation of such technology.

In a dual-phase LArTPC, particle tracking is achieved by utilising the ionisation electron signal that results from a particle's passage through the detector's liquid phase. A THick Gas Electron Multiplier (THGEM) \cite{THGEMReview_Breskin}, positioned in the gas phase, amplifies this charge signal, and it is then typically collected on a segmented anode plane.

However, the THGEM amplification process also produces secondary scintillation light - typically hundreds of photons per electron \cite{THGEMReadout, THGEMLightYield, THGEMReview_Buzulu}. This light can then be easily detected by modern optical imaging devices, which can be sensitive at the single-photon level. Improvements in spatial resolution may also be realised depending on the optical device(s) being used, as well as considerable simplification of detector design, construction and operation - leading to significant reductions in costs.
\\
\\
Optical readout of secondary scintillation light has previously been demonstrated by various groups using a number of different devices: Geiger-mode avalanche photodiodes \cite{BondarGAPD}, silicon photomultipliers \cite{THGEMReadout}, CCD cameras \cite{ccdargon} and EMCCD cameras \cite{emccdargon}. The ARIADNE LArTPC itself has been successfully operated using four EMCCDs \cite{ARIADNE_TDR}.

This paper will describe and discuss the optical readout of ARIADNE using a Timepix3-based camera. Such a device was previously used by the authors to demonstrate optical readout of a TPC containing low pressure CF$_4$ \cite{TPX3Paper}, and the results presented here represent a direct continuation of those studies.
\\
\\
In the following sections, an overview of the ARIADNE detector is first given, followed by a more detailed description of the Timepix3 camera hardware (referred to as a whole as ``TPX3Cam'' hereafter) and operational configuration. A selection of optically-imaged particle tracks is then presented, along with discussions on the electron lifetime correction and the results of detector calibration using through-going cosmic muons. Finally, a study of the energy deposition of stopping muons is presented.
 
\section{The ARIADNE Detector}

\subsection{Description and Detection Principle}
\begin{figure}[ht]
\centering
\includegraphics[width=0.55\textwidth]{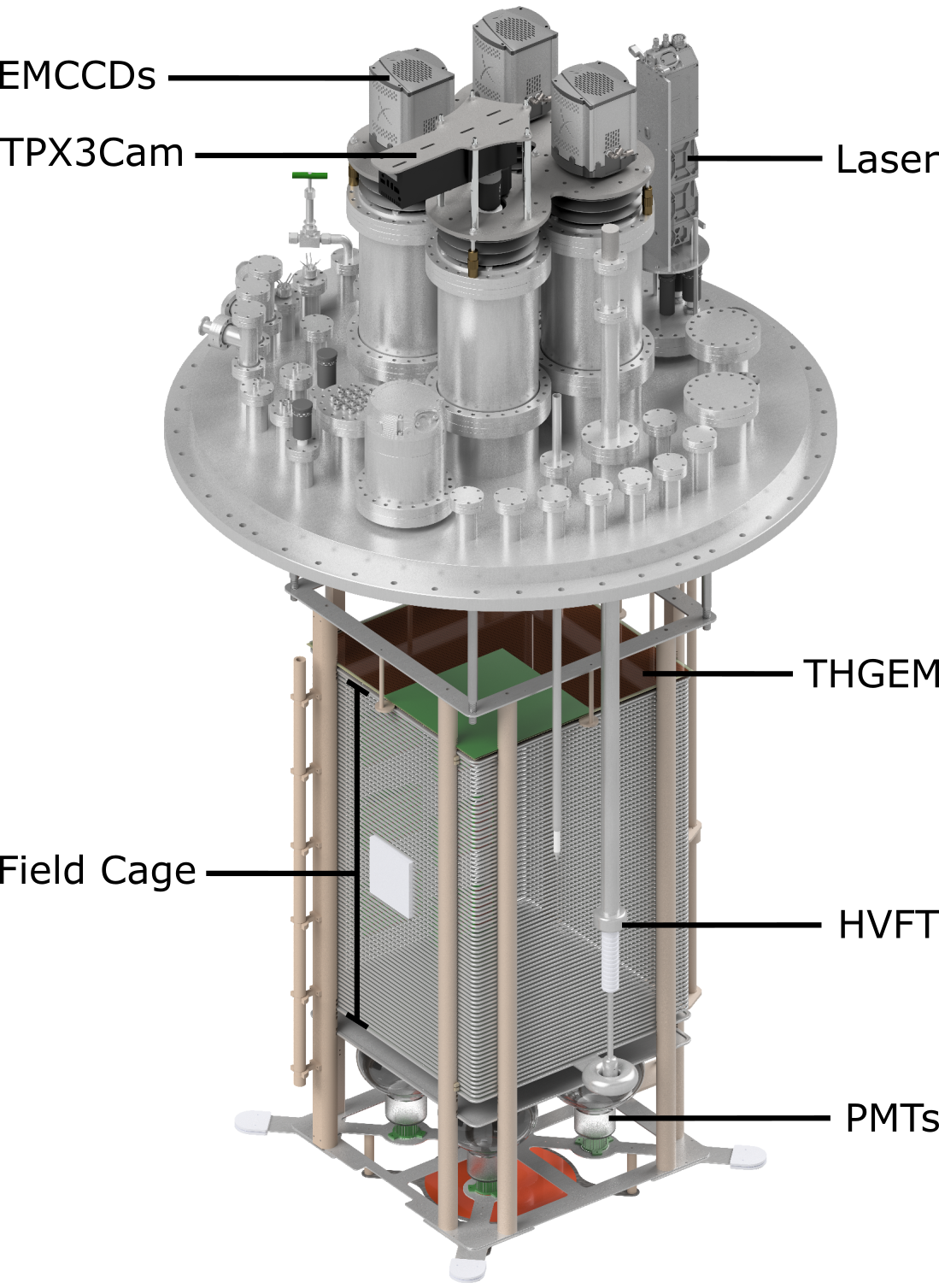}
\caption{A CAD model of the ARIADNE detector, with major components highlighted. The reflector panels have been omitted for clarity. Alongside the TPX3Cam are 3 EMCCDs, but these are shown only for illustrative purposes - their operation and results have previously been discussed in \cite{ARIADNE_TDR} and are not elaborated in this paper. The green region indicates the 32 $\times$ 32~cm field of view of the TPX3Cam.}
\label{fig:ARIADNEDetector}
\end{figure}

The following is a brief summary of the ARIADNE detector and detection principle. A more detailed description can be found in \cite{ARIADNE_TDR}.
\\
\\
Figure~\ref{fig:ARIADNEDetector} shows an overview of the ARIADNE detector, with major components highlighted.

At the centre of the detector is the 54 $\times$ 54 $\times$ 80~cm ($x$, $y$, $z$) field cage. This is constructed from 79 stainless steel field-shaping rings, with 1~cm centre-to-centre spacing and supported by VICTREX PEEK 450G rods. A cathode grid, capable of being biased up to 80~kV, is located below the field-shaping rings, and an extraction grid is positioned above them. Together, the grids and the field-shaping rings can produce a drift field of up to 1~kV/cm within the TPC. The field cage is surrounded on all four sides by reflector panels, consisting of 3M\textsuperscript{TM} Enhanced Specular Reflector Vikuiti\textsuperscript{TM} polyester foil coated with Tetraphenyl Butadiene (TPB).

A THick Gas Electron Multiplier (THGEM) is placed 11~mm above the extraction grid - this gap forms the extraction region, which is nominally biased with an extraction field of 3~kV/cm. The THGEM, manufactured by the CERN PCB Workshop, has an active area of 53 $\times$ 53~cm and a thickness of 1~mm. The holes populating the active area are arranged in a hexagonal array, with a pitch of 800~$\mu$m, and each individual hole having a diameter of 500~$\mu$m and an additional 50~$\mu$m dielectric rim. The Vacuum Ultraviolet (VUV) light produced in the THGEM holes is centred on a wavelength of 126 $\pm$ 7.8~nm (FWHM) \cite{larScintillation} - below the spectral sensitivity of the TPX3Cam readout system installed on ARIADNE. A glass plate coated with TPB is therefore positioned 2~mm above the top surface of the THGEM. This converts the VUV light to a wavelength of 420~nm, which is within the high quantum efficiency region of the TPX3Cam (discussed further in Section~\ref{subsec:TPX3Cam_Descr}).

Below the cathode grid are four TPB-coated Hamamatsu R5912-20-MOD 8~inch PMTs, which are used to provide trigger timing via the detection of prompt (denoted as ``S1'') scintillation light. They can also be used to provide complementary detection of the secondary (``S2'') scintillation light emitted from the THGEM.

\begin{figure}[ht]
\centering
\vspace{3mm}
\includegraphics[width=0.55\textwidth]{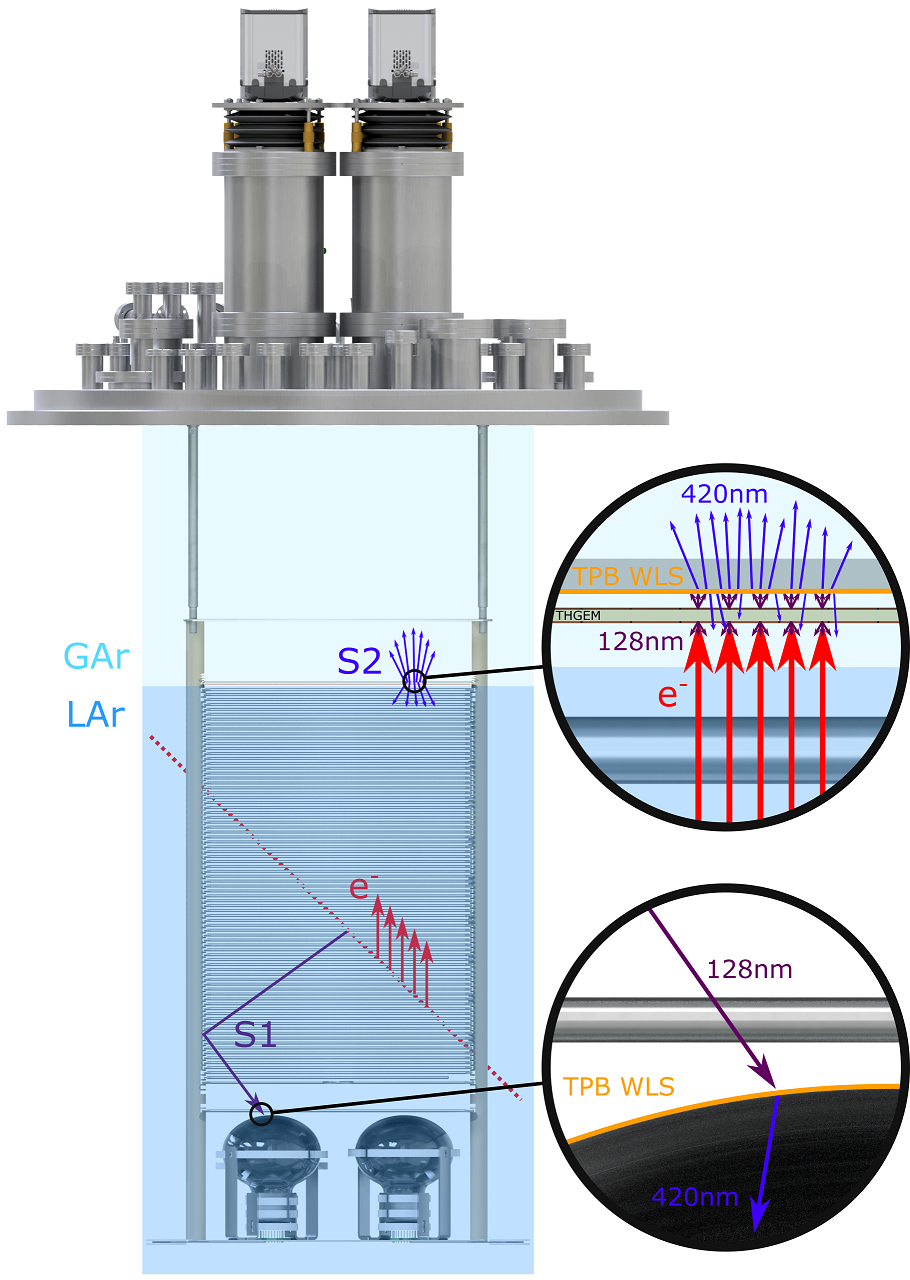}
\caption{The detection principle of ARIADNE. A charged particle enters the active LAr volume and produces a prompt (S1) signal, which is seen by the PMTs. Simultaneously, the argon atoms are also ionised, and the free electrons drift upwards in response to the drift field. At the liquid surface, the electrons are extracted into the gas phase, and further accelerated in the THGEM holes, producing secondary (S2) scintillation light which is detected by both the TPX3Cam and the PMTs.}
\label{fig:ARIADNEDetectionPrinciple}
\end{figure}

The chosen optical readout devices - currently a single TPX3Cam, and previously four EMCCDs - are positioned on optical viewports located on the top flange of the detector, looking down into the cryostat. For the studies discussed in this paper, the TPX3Cam was mounted on one of these viewports, with a field of view of 32 $\times$ 32~cm, covering slightly more than a quarter of the THGEM active area, as depicted in Figure~\ref{fig:ARIADNEDetector}.
\\
\\
A schematic of the detection principle is shown in Figure~\ref{fig:ARIADNEDetectionPrinciple}.

When a charged particle passing through the ARIADNE TPC interacts with the LAr, it both excites and ionises the argon atoms. This results in the emission of prompt (S1) scintillation light as the atoms de-excite, as well as the production of free electrons from the ionisation. The S1 signal is detected by the PMTs at the bottom of the TPC, and denotes the event start time. The free electrons are drifted upwards to the surface of the liquid phase by the drift field, which has a nominal gradient of 0.54~kV/cm (using cathode and extraction biases of -46.0~kV and -3.0~kV respectively).

At the surface of the liquid, the electrons experience a higher electric field (between the extraction grid and the bottom plane of the THGEM) and are extracted into the gas phase. Subsequently, they then enter the holes of the THGEM. The electric field across the THGEM (which has been varied between 16 and 30~kV/cm in these studies) is high enough to accelerate the electrons such that they gain sufficient energy to excite and ionise the gaseous argon atoms within the holes. The resulting de-excitations produce a secondary (S2) scintillation signal. This isotropically-emitted light is detected by both the PMTs below and the optical readout device(s) above the TPC.

\subsection{Fast Imaging using TPX3Cam Technology}
\label{subsec:TPX3Cam_Descr}
\begin{figure}[ht]
\centering
\includegraphics[width=0.80\textwidth]{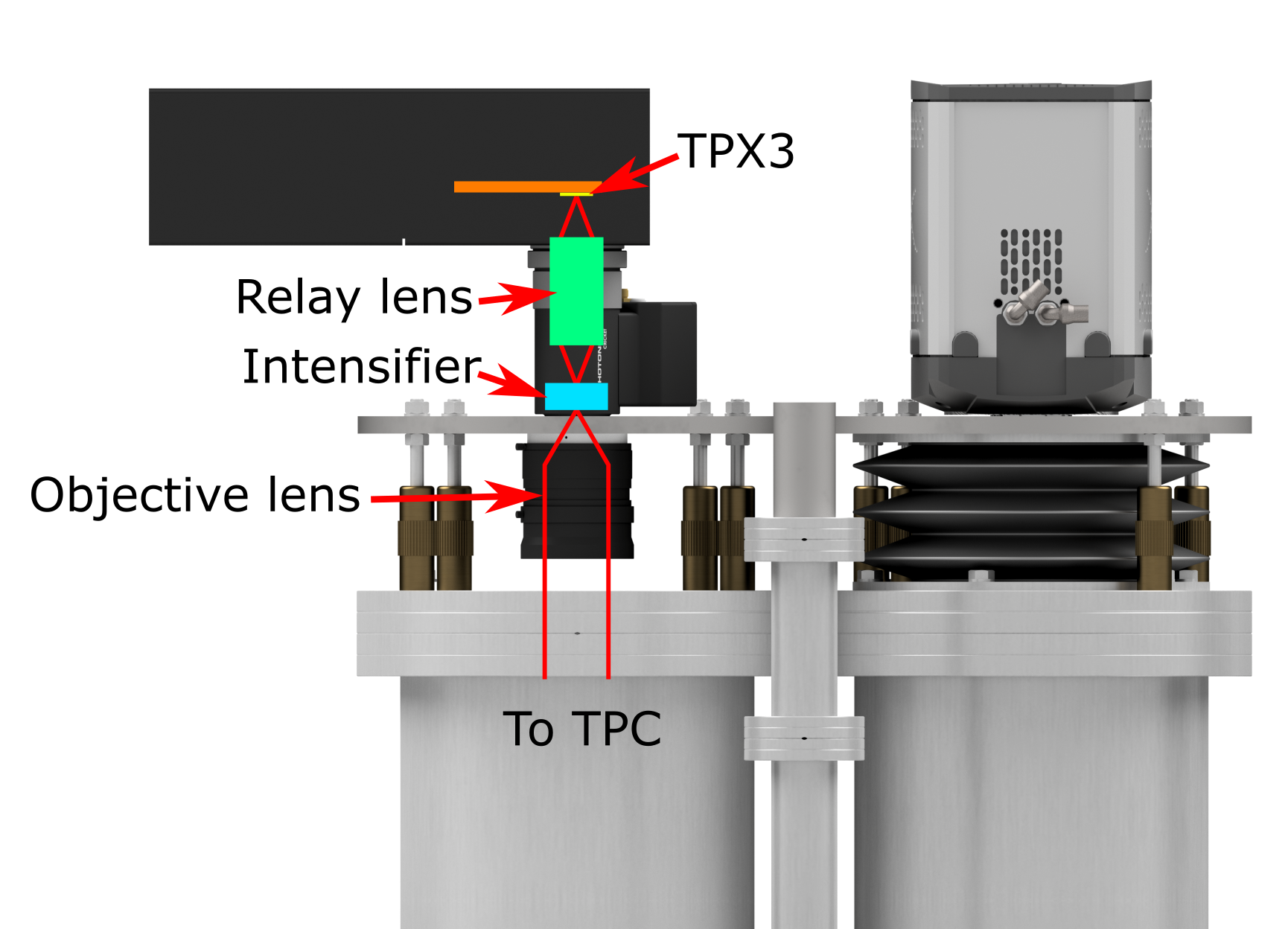}
\caption{The TPX3Cam detection principle. The light-tight bellow around the objective lens has been omitted for clarity. The objective lens focuses light generated in the THGEM holes onto a photocathode at the front of the image intensifier. The intensifier provides a photon gain of $\approx$ 10$^6$ photons/photon. A relay lens directs the intensified light signal onto a light-sensitive sensor which is bump-bonded to a TPX3 ASIC.}
\label{fig:TPX3CamDetectionPrinciple}
\end{figure}

As noted above, ARIADNE was previously operated using EMCCDs, which have readout rate limitations. To overcome these, a new camera based on TPX3 technology has been tested. The first successful demonstration of this technology was performed using the ARIADNE prototype vessel with low pressure CF$_4$ by the authors~\cite{TPX3Paper}. Following this demonstration, one EMCCD camera from ARIADNE was replaced with a TPX3Cam.
\\
\\
The operation principle of the TPX3Cam is shown in Figure~\ref{fig:TPX3CamDetectionPrinciple}. The assembly consists of an objective lens coupled to a Photonis Cricket image intensifier. The incident S2 light from the THGEM is converted to electrons using a photocathode, and these electrons are then multiplied through dual microchannel plates (MCPs). The amplified electron signal is then converted back to light using a P47 phosphor screen. The photons from the phosphor screen are focused, using a relay lens, onto a light-sensitive silicon sensor, which is bump bonded to a TPX3 ASIC (a 256 $\times$ 256 pixel array with 55~$\mu$m pixel pitch) \cite{TPX3ASICPaper, TPX3ASICPaper2}. The photons are converted into electron-hole pairs in the sensor, and the TPX3 ASIC then accumulates these electrons.

\begin{figure}[ht]
\centering
\includegraphics[width=0.75\textwidth]{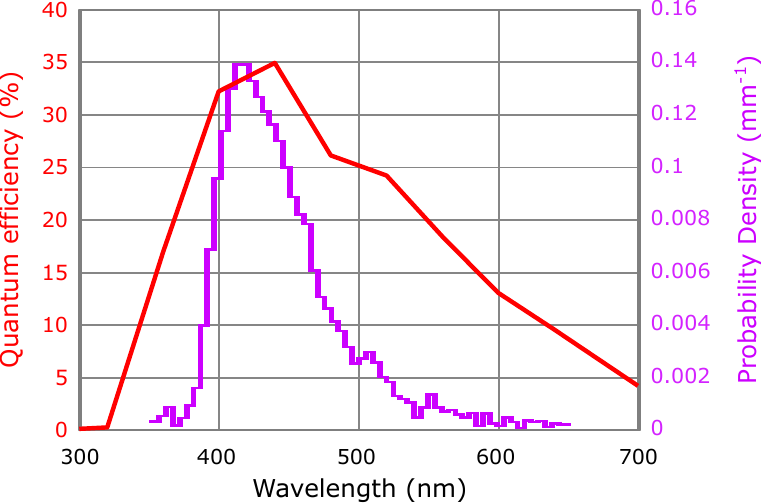}
\caption{(Red) The quantum efficiency spectrum of the Photonis image intensifier used in this work, as measured by the manufacturer. (Purple) The TPB emission spectrum, adapted from \cite{TPBSpectrum}.}
\label{fig:IntensifierQE}
\end{figure}

The quantum efficiency of the silicon sensor is approximately 90$\%$ at the peak emission wavelength (420~nm) of the P47 phosphor screen \cite{TPX3CamQEPaper, TPX3CamQEPaper2, TPX3CamQEPaper3}. If the total accumulated charge exceeds a user-defined threshold (a typical minimum operating threshold is $\approx$ 500~e$^{-}$), the TPX3 chip registers a hit. The start time of this hit, known as the Time of Arrival (ToA), is recorded with 1.6~ns resolution. The accumulated charge is then discharged, at a constant rate, until the charge has dropped below the predefined threshold. The length of time for which the charge remained over threshold is known as the Time over Threshold (ToT). This value is  proportional to the number of photons incident on the pixel of the silicon sensor (similar to intensity in an EMCCD). The ToT is recorded with 10-bit resolution, giving a dynamic range of 1024 ADU (analog-to-digital units).
\\
\\
The image intensifier used in this work is a Photonis Cricket. This unit combines an image intensifier tube with a relay lens in a single housing with C-mount interfaces for both objective lens and installation onto a camera. The intensifier is configured with a `Hi-QE Green' photocathode, with 30-35$\%$ quantum efficiency at 420~nm, as measured by Photonis after manufacture. Figure \ref{fig:IntensifierQE} compares the quantum efficiency of the image intensifier to the emission spectrum of TPB. The intensifier tube has two microchannel plates (MCPs) arranged in Chevron configuration, and a P47 phosphor screen to convert the amplified electron signal from the MCP back into photons. The intensifier tube has a photon gain of 1.1 $\times$ 10$^6$ photons/photon, as measured at the factory after manufacture.\\

\noindent Readout of the TPX3 ASIC is facilitated by a SPIDR readout board \cite{SPIDRPaper}, which is itself incorporated within the TPX3Cam. The SPIDR readout board includes both 1 and 10 Gigabit Ethernet interfaces for data transfer between the TPX3Cam and a DAQ computer. The data stream from the camera consists of a series of 64-bit `hit' packets, each containing 4 pieces of information: the column and row of the hit pixel (related to the $x$ and $y$ position within the TPC), ToA (related to the $z$ position) and ToT. These packets are produced in a data driven / sparse readout format (i.e. hits are generated only for those pixels which go over threshold). The maximum readout rate is approximately 80~MHits/s. These properties make TPX3 well suited for optical readout, due to the high readout rate, natively 3D raw data, and low storage requirements due to zero suppression.

\begin{figure}[ht]
\centering
\includegraphics[width=0.95\textwidth]{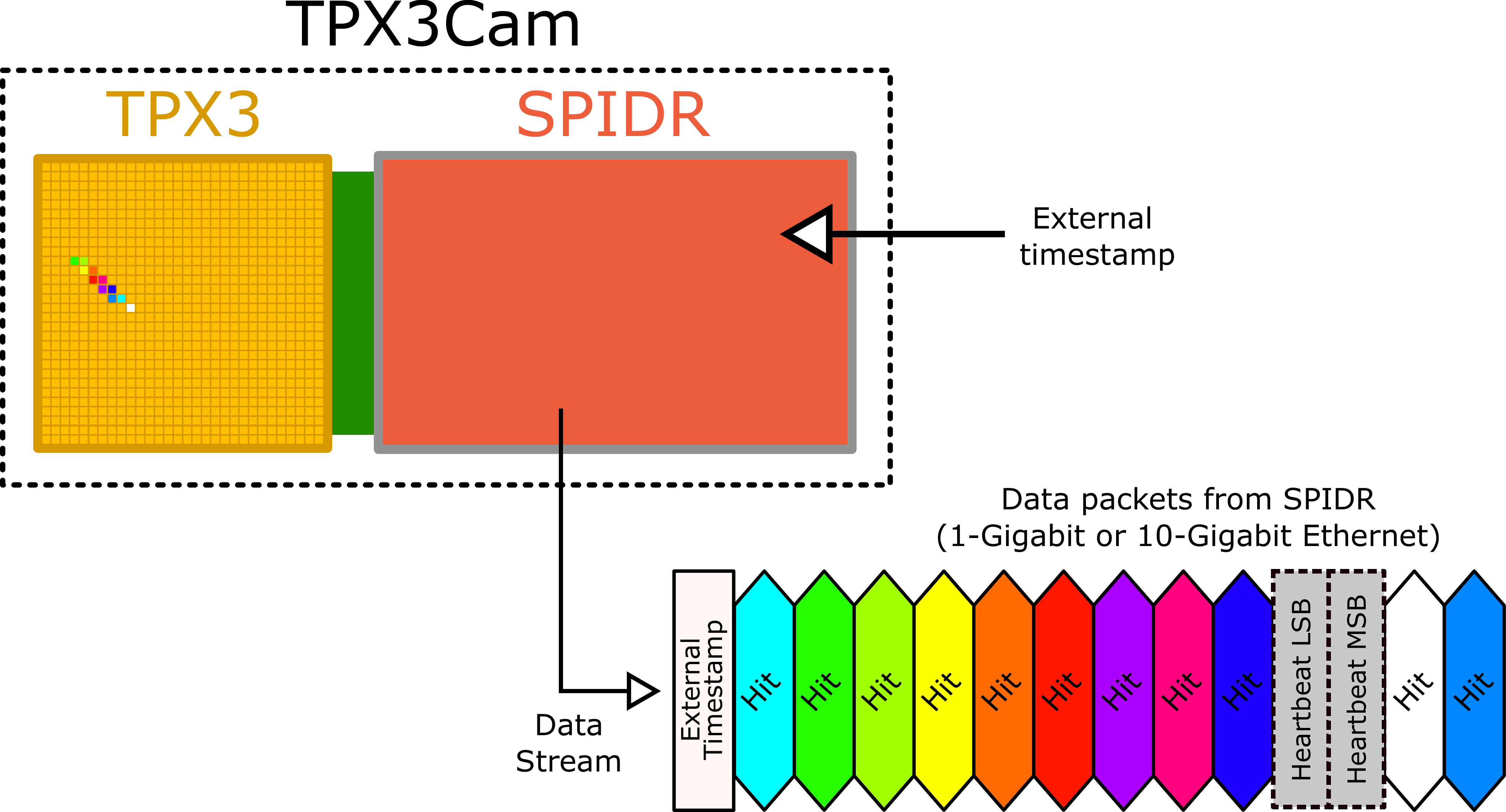}
\caption{General overview of the operation of the SPIDR readout board. A continuous stream of data packets is generated by the SPIDR readout board, encoding all of the information taken during operation, including the external timestamps. The packets are transferred to storage via a 1 Gigabit Ethernet connection, and are then parsed and analysed offline. A more detailed description can be found in \cite{RobertsThesis}.}
\label{fig:DataDrivenReadout}
\vspace{10mm}
\end{figure}

The SPIDR readout board also includes the option to insert timestamps into the data stream. When an external signal is received by SPIDR, a timestamp packet is generated and inserted into the datastream, which records the time at which the signal was received with 200~ps resolution. Figure~\ref{fig:DataDrivenReadout} illustrates the readout principle.

\subsection{Operational Configuration}
\label{subsec:OpsConfig}
Table~\ref{tab:OpConfig} shows the various parameters that were used during operation of the ARIADNE detector, arranged by component. (The TPX3Cam configuration was previously discussed in Section \ref{subsec:TPX3Cam_Descr} above.)

\begin{table}[h!]
    \centering
    \begin{tabularx}{\textwidth}{|p{3.0cm}|p{1.75cm}|p{9.1cm}|}
        \cline{1-3}
        \textbf{Parameter} & \textbf{Value} & \textbf{Comments} \\
        \cline{1-3}
        \cline{1-3}
        \textbf{PMTs} & & \\
        \cline{1-3}
        Biases & 970V \newline 1100V \newline 1300V \newline 950V & Different voltages used so as to equalise the individual PMT responses \\
        \cline{1-3}
        \cline{1-3}
        \textbf{Field Cage Biases} & & \\
        \cline{1-3}
        Cathode \newline Extraction Grid & -46.0~kV \newline -3.0~kV & Drift field of 0.54~kV/cm \\
        \cline{1-3}
        THGEM Top \newline THGEM Bottom & +1.945~kV \newline -1.000~kV & Nominal THGEM field of 29.45~kV/cm  (varied for the THGEM characterisation study in Section~\ref{subsec:THGEMCharacterisation})\\
        \cline{1-3}
        \cline{1-3}
        \textbf{Other} & & \\
        \cline{1-3}
        Cryostat pressure & 1.040~bar & \\
        \cline{1-3}
    \end{tabularx}
    \caption{Operational parameters for the ARIADNE detector.}
    \label{tab:OpConfig}
\end{table}

\noindent For this work, the TPX3Cam is connected to a DAQ computer using 1 Gigabit Ethernet, and a single PMT (directly underneath the TPX3Cam's field of view, previously shown by the green area in Figure~\ref{fig:ARIADNEDetector}) was connected to a CAEN V1720 digitiser. A trigger signal was generated from the digitiser when the PMT signal exceeded a predefined threshold. This threshold was set empirically so as to select higher energy cosmic events, and the resulting PMT trigger rate varied between 50 and 100~Hz. The trigger signal was then passed to the TPX3Cam's external trigger input, acting as an external timestamp. This timestamp - denoting the S1 signal detection time as seen by the PMT - is therefore equivalent to the event start time. For the purposes of all subsequent analyses, an event is defined as the collection of all hits with a ToA within a 600~$\mu$s window following each external timestamp.

\section{Cosmic Muons Analysis}

\subsection{Spatial Conversion Factors and Drift Velocity}
\label{subsec:SpatialCalibs}
As noted in Section~\ref{subsec:TPX3Cam_Descr}, the information packet from each hit pixel contains the column and row numbers corresponding to the pixel's position on the sensor, as well as the ToA and the ToT. The first three of these quantities are respectively related to the physical $x$, $y$ and $z$ coordinates of the corresponding energy deposition within the detector. The conversion factors from raw to physical distance units are determined as follows.
\\
\\
Converting the pixel column and row to $x$ and $y$ coordinates involves understanding the positioning of the TPX3Cam's field of view with respect to the detector.

The ARIADNE design nominally sets the distances from the central axis of each optical viewport to the nearest edges of the THGEM active area as 135~mm (in both $x$ and $y$). An integrated TPX3Cam image is used to determine the number of pixels that this distance corresponds to, and equating the results of this to the design specifications thus yields the $x$ and $y$ conversion factors:

\begin{equation}
  \begin{array}{l}
    x~(mm) = 1.26 \times \textrm{hit column number} \\
    y~(mm) = 1.26 \times \textrm{hit row number}
  \end{array}
\label{eq:xySpatialCalibrations}
\end{equation}
\\
\noindent The ToA of each hit is related to the corresponding $z$ coordinate via the mean drift velocity, $v_d$, of electrons within the TPC. This can be calculated by considering the maximum possible drift time, $t_d$, of electrons in the TPC - i.e. those that begin drifting at the cathode.

\begin{figure}[ht]
\includegraphics[width=0.5\textwidth]{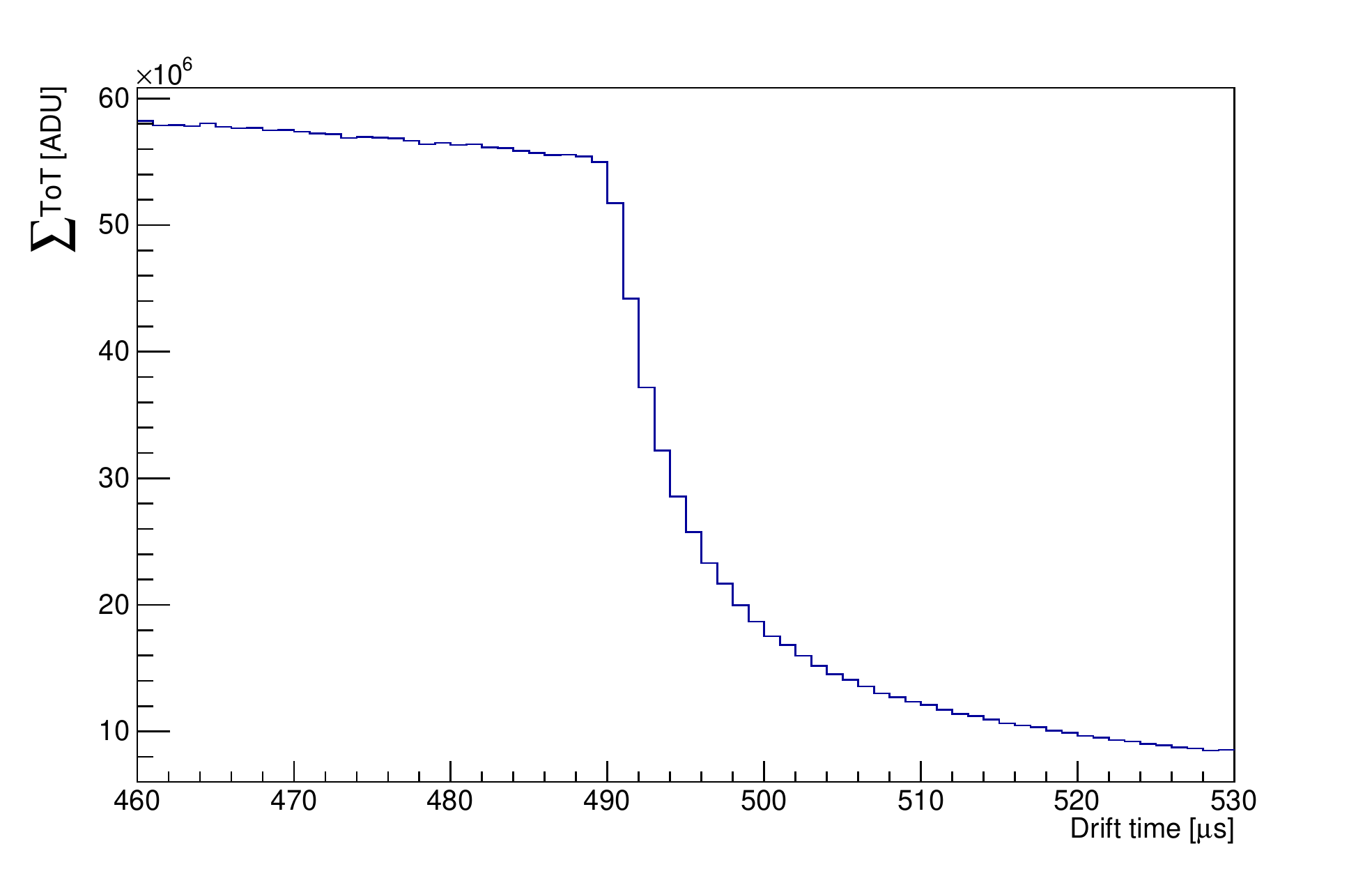}
\includegraphics[width=0.5\textwidth]{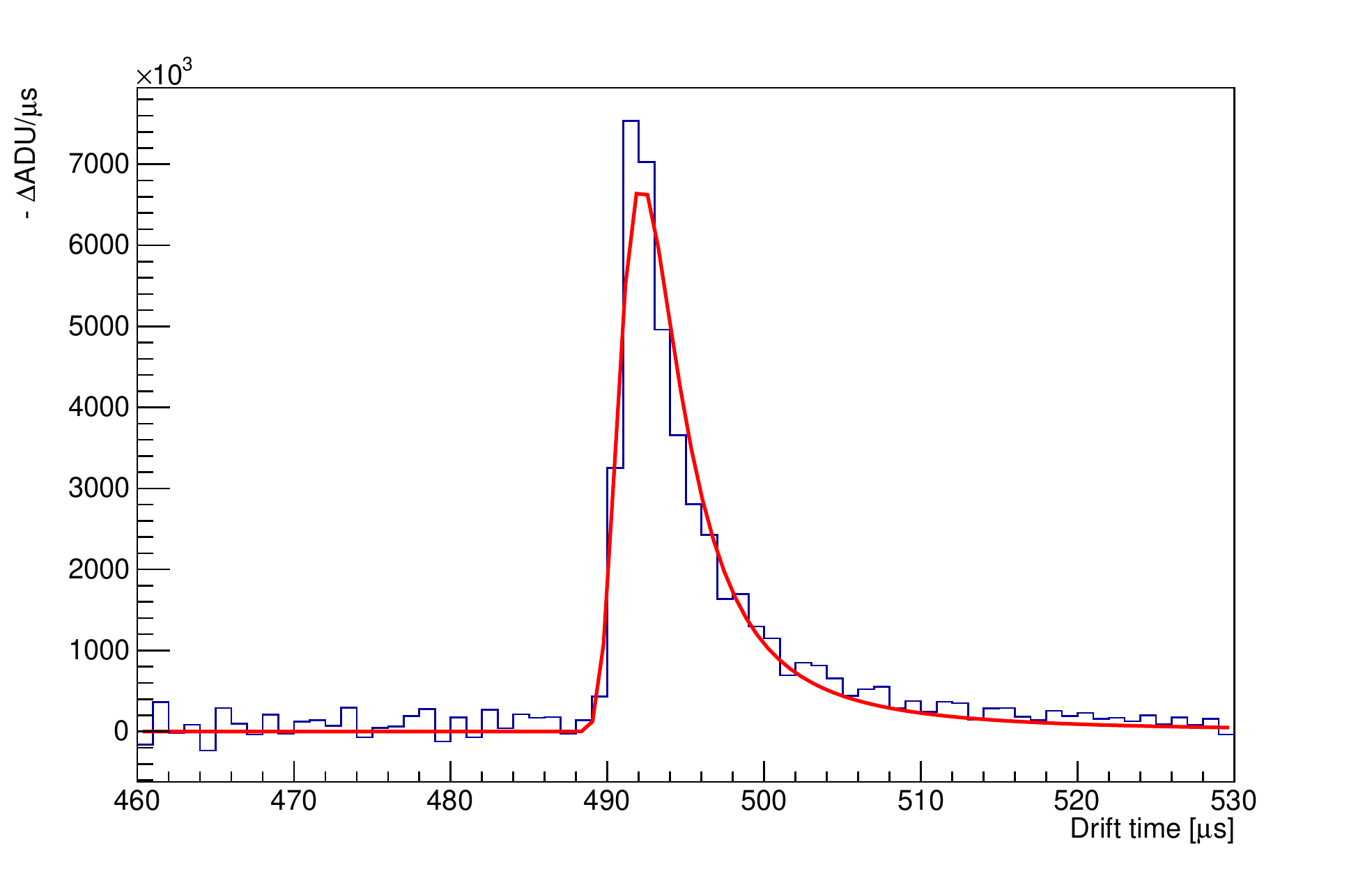}
\caption{(Left) The summed ToT of hits as a function of the drift time $t_d$, for a small range around the time approximately corresponding to the bottom of the TPC. A sharp drop is observed, indicating the cathode. (Right) The negative gradient of the summed ToT distribution. The steep drop now takes the form of a peak, to which a Landau function has been fitted.}
\label{fig:driftTimeDelta}
\end{figure}

The distribution of the summed ToT as a function of $t_d$ (which is equivalent to the hit time relative to the event start time for the purposes of this study) is found to be consistent with hits that are distributed throughout the entire detector volume. A sharp drop in the distribution, corresponding to the bottom of the TPC, is found around 500~$\mu$s, as shown in Figure~\ref{fig:driftTimeDelta} (left). The extended tail of the distribution is believed to be caused by event pileup. Figure~\ref{fig:driftTimeDelta} (right) shows the negative gradient of the summed ToT distribution in the same region. A Landau function has been empirically found to describe the distribution shape very well, and is therefore fitted to the gradient distribution. The most-probable-value (``MPV'') of the Landau denotes the point of largest gradient, i.e. the steepest drop in summed ToT, which is assumed to occur at the cathode itself. The MPV is therefore assumed to be equal to $t_d$. Although the statistical error on this is then given by the error on the MPV, the width parameter of the Landau has been used to better represent the combined statistical and systematic uncertainties.

\begin{figure}[ht]
\centering
\includegraphics[width=0.8\textwidth]{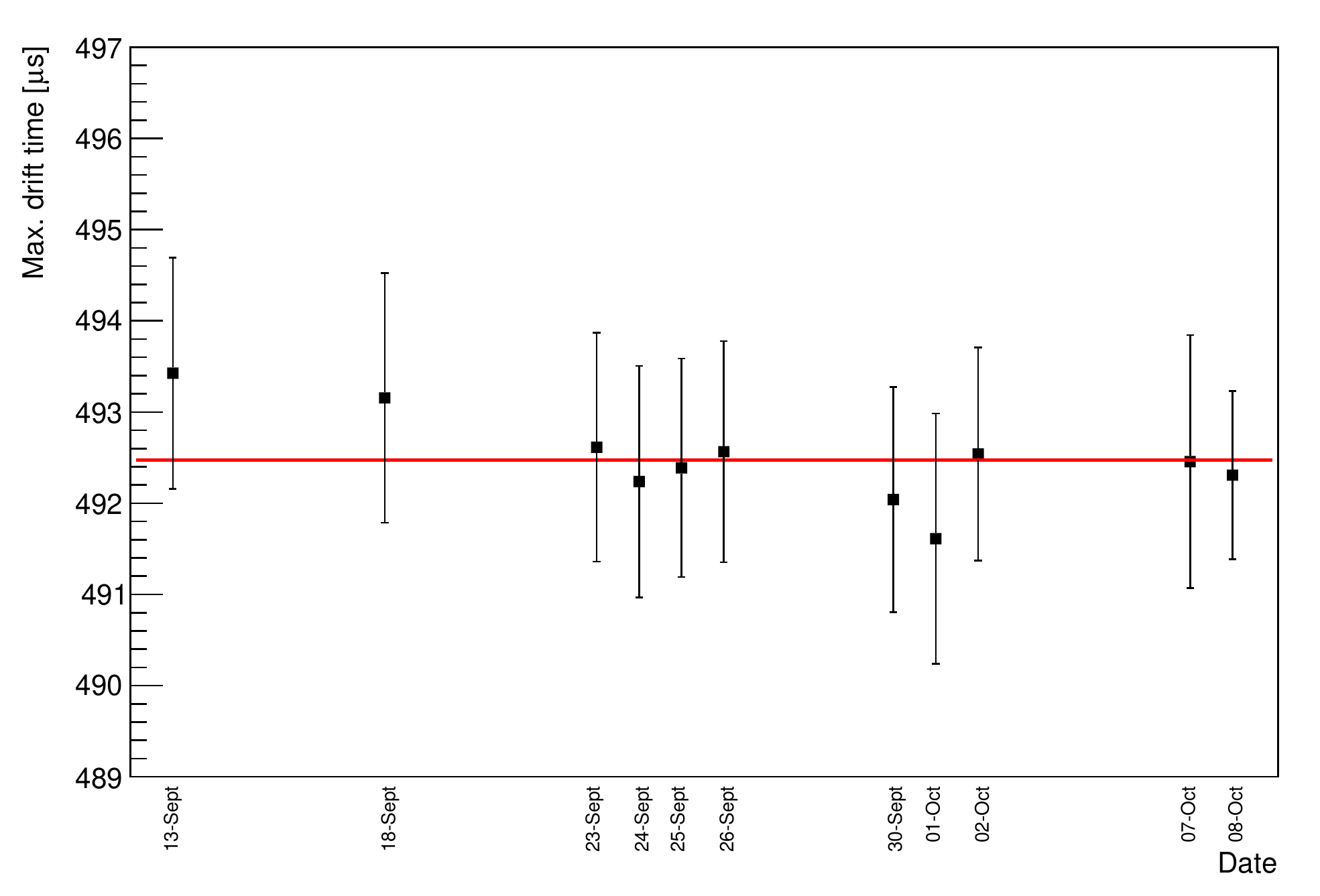}
\caption{The maximum possible drift time $t_d$, as calculated for a series of datasets taken over a period of several weeks. The error bars correspond to the width parameter of each dataset's fitted Landau function. The mean value across all datasets is shown in red.}
\label{fig:maxDriftTimeProgression}
\end{figure}

\noindent Figure~\ref{fig:maxDriftTimeProgression} shows the maximum possible $t_d$ across a number of datasets taken over a period of several weeks. It can be seen that it is very stable, only fluctuating by a few $\mu$s over this period. The statistical mean of these results - found to be $492.5 \pm 0.4~\mu$s - can therefore be taken as a reliably representative value.
\\
\\
Along with the maximum possible $t_d$, the maximum drift length - i.e. the total height of the TPC - is also required in order to calculate $v_d$. Nominally, this is 800~mm as previously described, however this does not account for shrinkage due to the cryogenic temperature of LAr. The contraction rate of VICTREX PEEK 450G is $50 \pm 10$~ppm/K \cite{PEEKProperties}. A conservative estimate of the uncertainty has been used, as the exact rate depends on the direction of contraction with respect to the material flow. For the ARIADNE TPC under a temperature differential of ( 292 $\rightarrow$ 87 = ) 205~K, this corresponds to a reduction of 8.2~mm, giving a cryogenic TPC height of $791.8 \pm 2.6$~mm (with uncertainties on the nominal TPC height and the temperature differential also accounted for).

Combining this with the previously found mean maximum possible $t_d$ gives:

\begin{equation}
v_d = 1.608 \pm 0.005~mm/\mu\textrm{s}
\label{eq:driftVelocity}
\end{equation}
\\
This value is in good agreement with other experiments \cite{DriftVelocity1, DriftVelocity2}. The conversion from hit ToA to the physical $z$ coordinate is then simply given by:

\begin{equation}
  z~(mm) = v_d \times \textrm{ToA}~(\mu\textrm{s})
\label{eq:zSpatialCalibration}
\end{equation}
\\
The three spatial conversions described above have been applied to all results discussed henceforth, unless otherwise stated.

\subsection{Gallery}
\label{subsec:Gallery}
\begin{figure}[ht]
\centering
\includegraphics[width=0.8\textwidth]{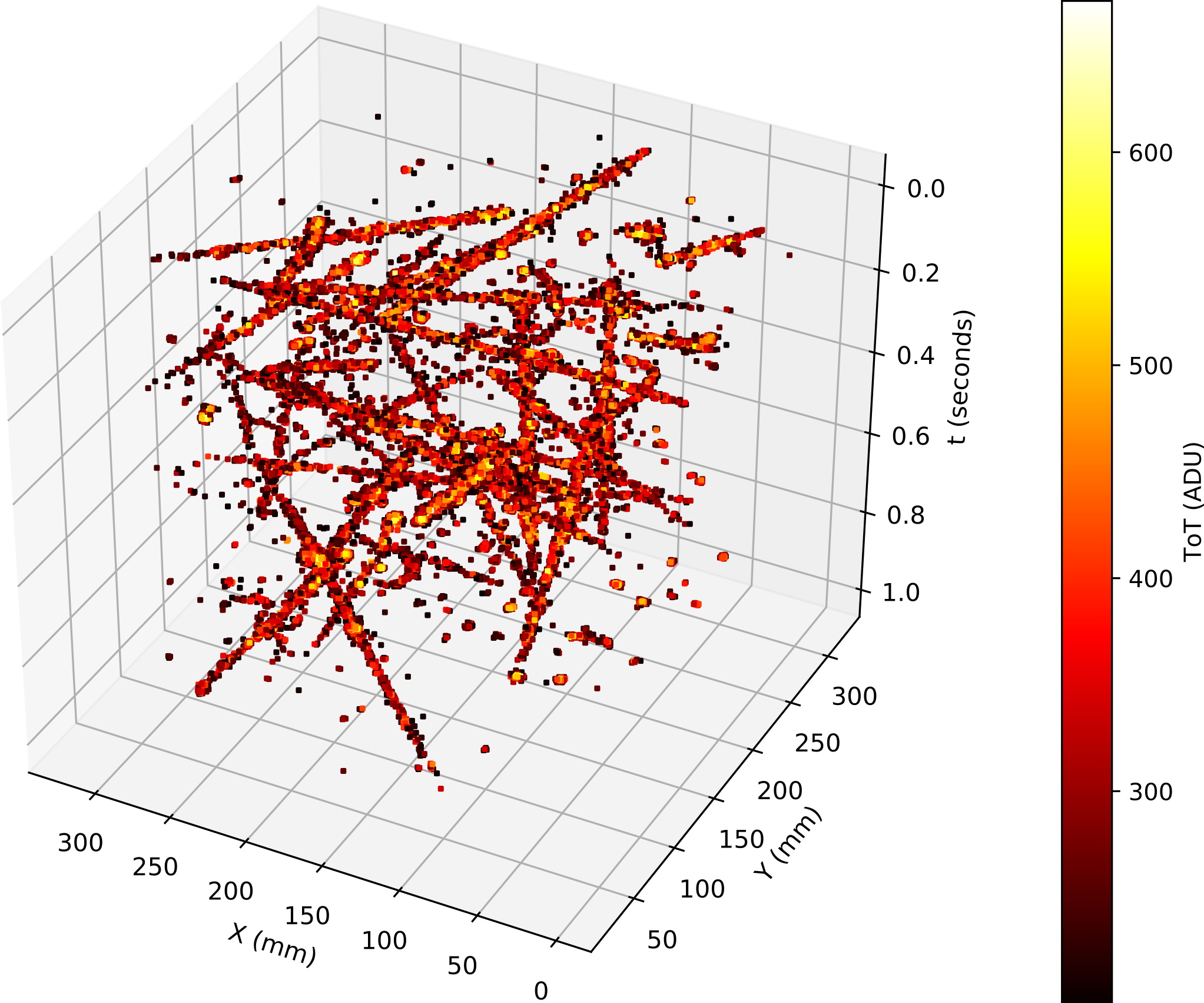}
\caption{A 1~second long slice of the TPX3Cam's continuous output data stream, showing numerous LAr interactions of cosmic muons, and corresponding to an effective drift length of 1.608~km. The colour bar indicates the individual hit ToT values in ADU.}
\label{fig:continuousStream}
\end{figure}

\begin{figure}[ht!]
\centering
\includegraphics[width=0.49\textwidth]{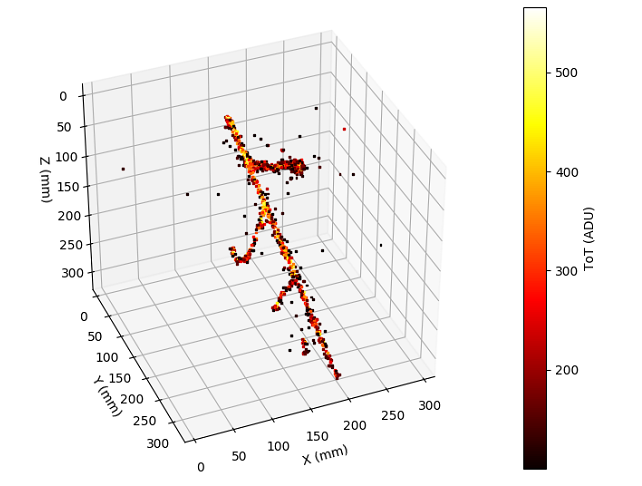}
\includegraphics[width=0.49\textwidth]{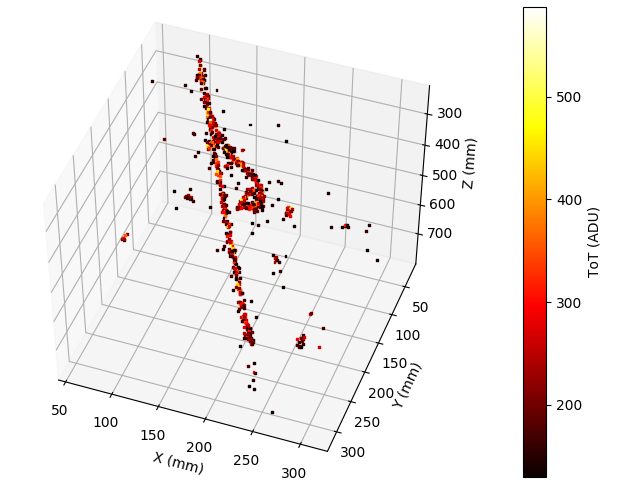}
\includegraphics[width=0.49\textwidth]{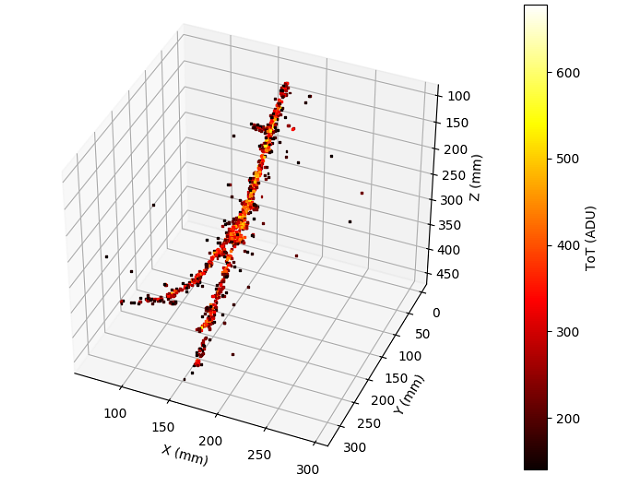}
\includegraphics[width=0.49\textwidth]{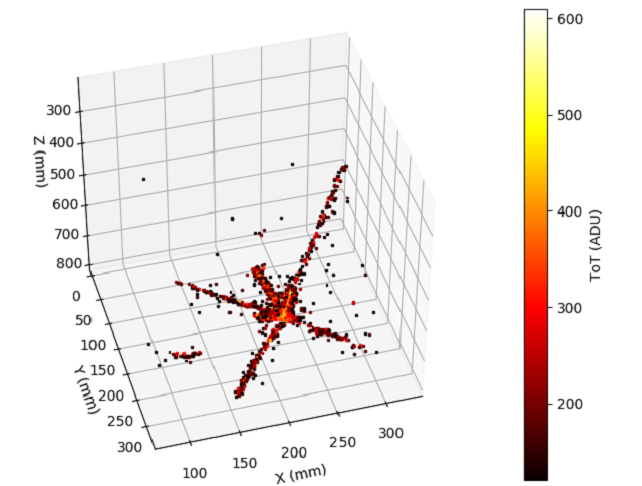}
\includegraphics[width=0.49\textwidth]{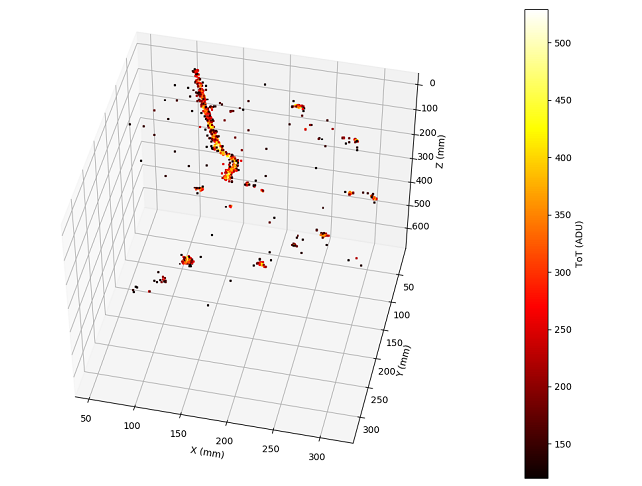}
\includegraphics[width=0.49\textwidth]{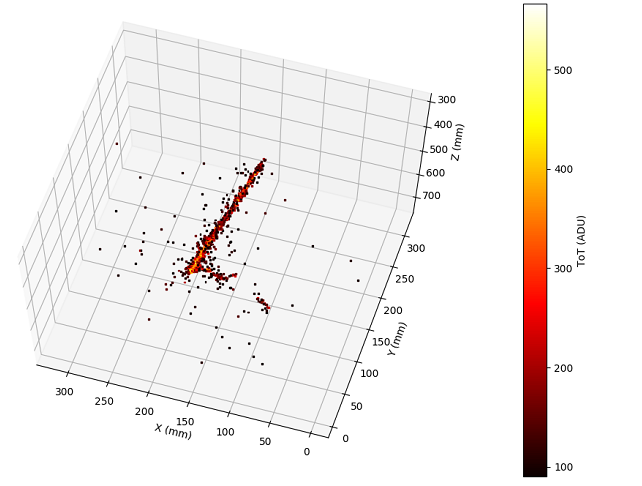}
\caption{A selection of cosmic muon-initiated LAr interactions imaged by the ARIADNE TPX3Cam: long muon tracks with one or more deltas rays (top row and middle-left), an anti-proton candidate (middle-right) and stopping muons with associated Michel electrons (bottom row). The colour bars indicate the hit ToT in ADU.}
\label{fig:cosmicsGallery}
\end{figure}

As discussed previously in Section~\ref{subsec:TPX3Cam_Descr}, the TPX3Cam outputs a continuous stream of hits, and a 1~second slice of this stream is shown in Figure~\ref{fig:continuousStream}. Using the drift velocity given in Equation~\ref{eq:driftVelocity}, this corresponds to an effective drift length of 1.608~km. It is obvious that a large number of cosmic particles have passed through the active LAr volume during this time window, but even at this scale, many of the individual tracks are quite distinguishable, due to the zero suppression of the raw data.
\\
\\
\noindent Figure~\ref{fig:cosmicsGallery} shows a selection of single-particle LAr interactions as imaged by the TPX3Cam. A variety of different particle types was observed: through-going muons with relatively simple geometries, muons which came to a stop in the detector (``stopping muons'') to produce an observable Michel electron, and a small number of possible anti-proton annihilations, which produce a characteristic event topology consisting of a vertex accompanied by multiple outgoing daughter particle tracks.

\subsection{Electron Lifetime Correction}
The amount of light observed by the TPX3Cam is partly dependent on the LAr purity. This in turn is closely related to the electron lifetime, denoted by $\tau$ - a measure of the relative fraction of drift electrons that are lost within the LAr due to attachment to impurities. Measurement of $\tau$ is therefore an excellent way to monitor changes in the LAr purity over time.
\\
\\
The value of $\tau$ is related to the energy deposition per unit length, $\frac{dE}{dX}$, and the drift time, $t_d$, by:

\begin{equation}
\frac{dE}{dX} = A e^{- \frac{t_d}{\tau}}
\label{eq:lifetimeExponential}
\end{equation}
\\
where $A$ is a constant. Axially through-going cosmic muons are therefore ideal for measuring $\tau$, as they produce long vertical tracks with constant energy deposition per unit length, and travel quite deep into the TPC, giving a large $t_d$ range.

\begin{figure}[ht]
\centering
\includegraphics[width=0.9\textwidth]{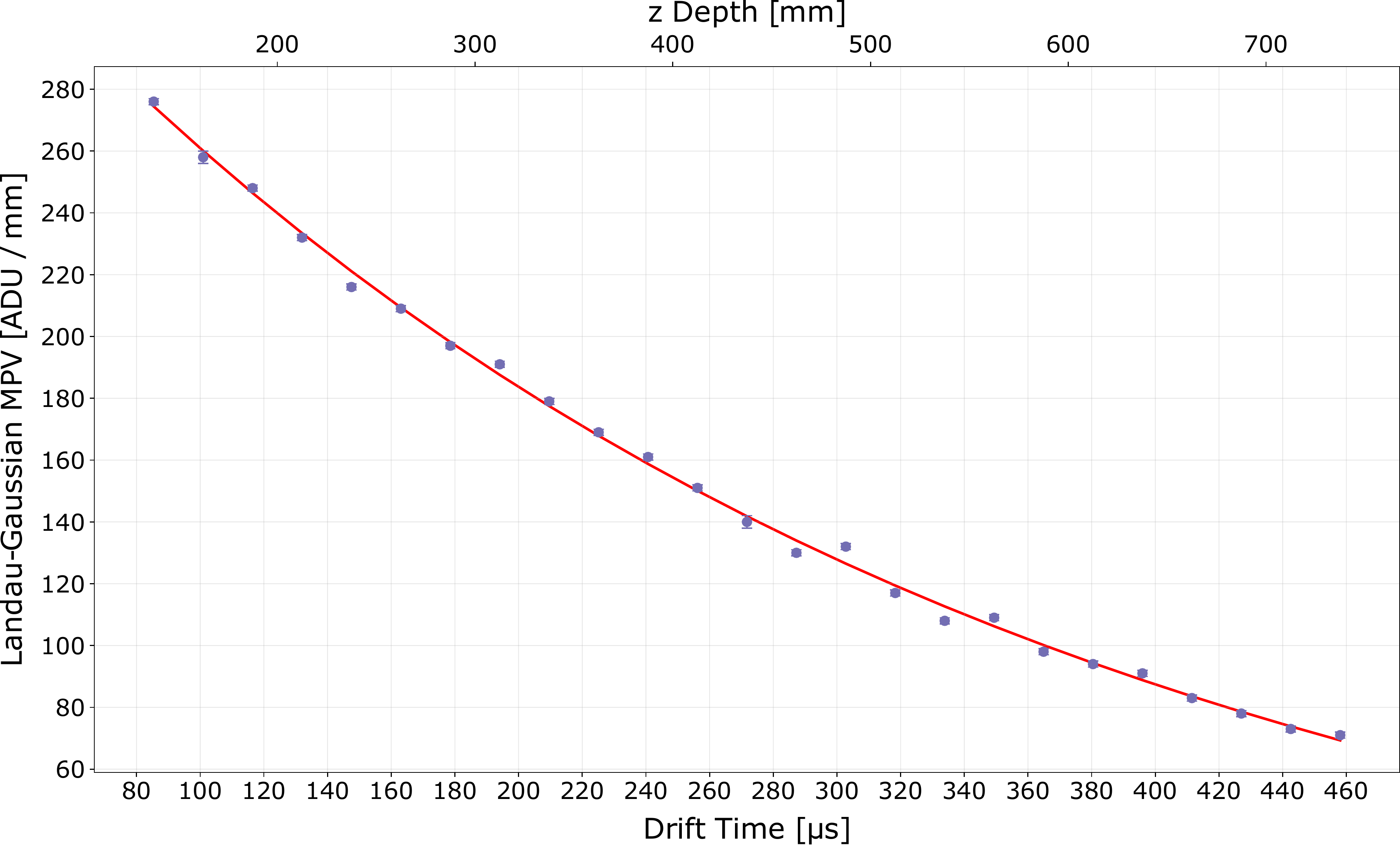}
\caption{The fitted Landau-Gaussian MPV of the $\frac{dI}{dX}$ per $t_d$ bin as a function of the (central) $t_d$ bin value (bottom $x$ axis). Also indicated (top $x$ axis) is the corresponding $z$-depth of each $t_d$ value. Error bars indicate only the statistical uncertainties. The MPV values have been fitted with an exponential function of the form given by Equation \ref{eq:lifetimeExponential} in order to determine the electron lifetime.}
\label{fig:lifetimeCorrection}
\end{figure}

The active LAr volume is first segmented into equal $z$-depth slices. The slice width was chosen to be 25~mm, corresponding to a drift time of $\approx$15.5~$\mu$s per slice, and giving 32 slices covering the entire height between the extraction and cathode grids. A single through-going cosmic muon will therefore pass through multiple $z$-slices. Within each slice that the muon occupies, a ToT summation is performed over all hits, and the 3D length of the muon track segment is also calculated. The ratio of these quantities is then the $\frac{dI}{dX}$ of this slice (where $dI$ corresponds to the summed ToT, measured in ADU and directly proportional to the observed energy deposition $dE$). When performed across a large number of muons, the $\frac{dI}{dX}$ values for the same $z$-slice form a distribution, which is fitted with a Landau-Gaussian convolution function. This has been chosen in order to account for both the characteristic Landau shape of the energy deposition distribution and the combined detector effects which tend to broaden it. The MPV of this Landau-Gaussian fit is used as the representative value of $\frac{dI}{dX}$ in this $z$-slice, and therefore at this corresponding $t_d$.
\\
\\
Figure~\ref{fig:lifetimeCorrection} shows the distribution of Landau-Gaussian MPVs as a function of the drift time $t_d$ for a single dataset of through-going cosmic muons. The $z$-depths corresponding to the $t_d$ scale are also shown, as is the fit of Equation \ref{eq:lifetimeExponential} to the MPV values. The error bars on each point indicates only the statistical uncertainties, equivalent to the error on the Landau-Gaussian MPV. It is expected that systematic uncertainties would be the dominant source of error, but a study of these has not been performed here.

\noindent It can be seen that the MPV values do indeed follow an exponential distribution, and this particular fit corresponds to an electron lifetime of $309 \pm 7$ (stat.)~$\mu$s for this dataset. Other TPX3Cam datasets have been analysed using this method, and their electron lifetimes values similarly calculated.
\\
\\
The calculated electron lifetime $\tau$ is used to ``correct'' the hit ToT values within each dataset, so as to compensate for the reduction in light intensity due to the loss of electrons produced lower down in the TPC. This correction takes the form:

\begin{equation}
ToT \rightarrow ToT \times e^\frac{z}{\tau v_d}
\label{eq:lifetimeCorrection}
\end{equation}
\\
where $v_d$ is the drift velocity found previously in Equation \ref{eq:driftVelocity}. This correction has been applied to the results discussed henceforth, unless otherwise stated.

Although ARIADNE has an internal purification and recirculation system (described in \cite{ARIADNE_TDR}), further purification was not possible during this operation due to saturation as the result of a prior leak. In future ARIADNE operation, a higher argon purity is anticipated to be achieved, removing the need to apply the electron lifetime correction. In general, high argon purity levels are now routinely achieved in liquid argon TPCs, as demonstrated for example in \cite{ProtoDUNESPResults}.

\subsection{THGEM Characterisation}
\label{subsec:THGEMCharacterisation}
The S2 light output of the THGEM as a function of the electric field across it was measured using the TPX3Cam.

\begin{figure}[ht]
\centering
\includegraphics[width=0.75\textwidth]{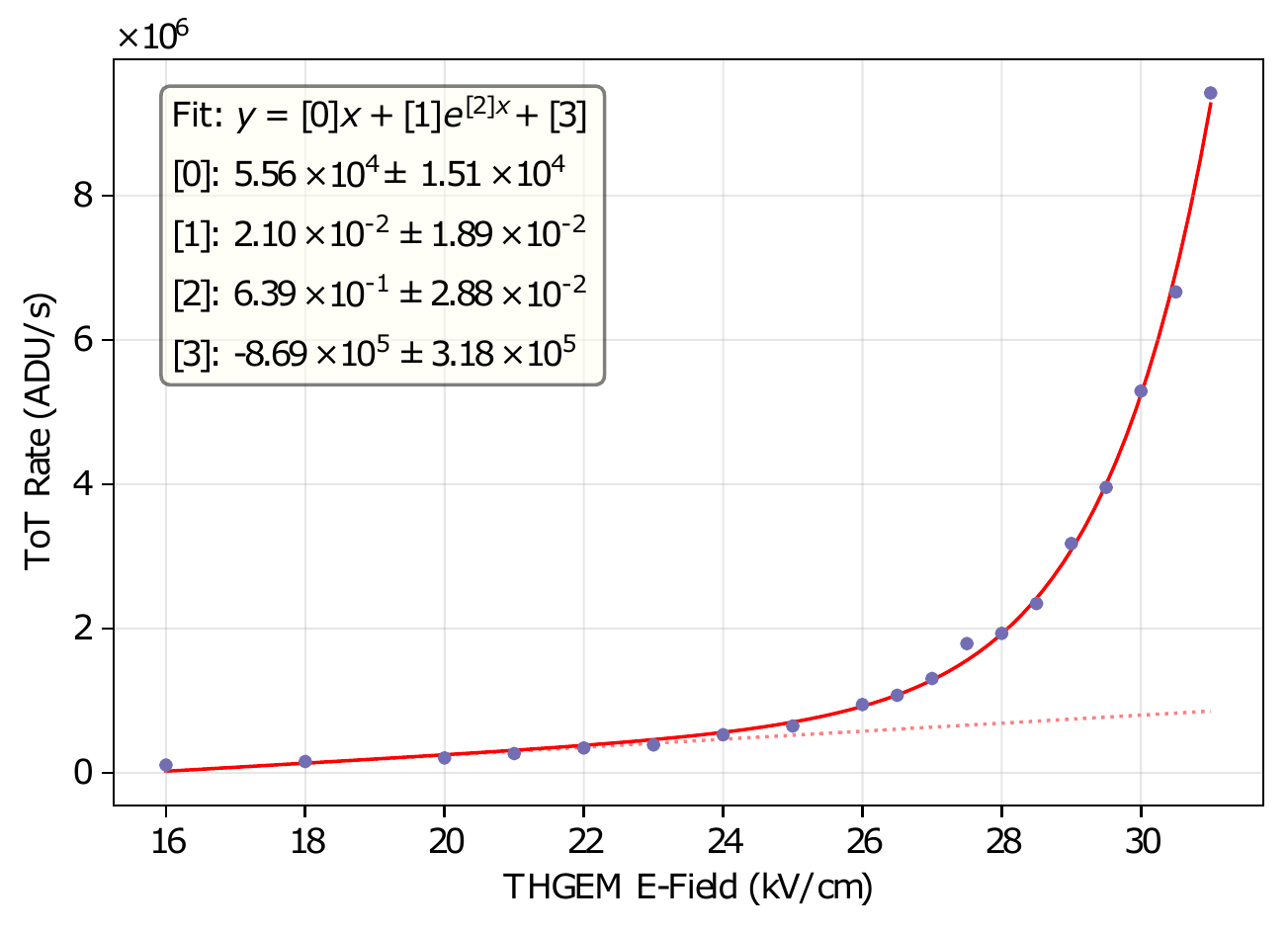}
\caption{The TPX3Cam ToT rate as a function of the electric field across the THGEM. A single function as given in the legend has been fitted to the data, and the linear and exponential regimes (discussed in the main text) can be observed. Error bars have been omitted as they are within the datapoint markers.}
\label{fig:THGEMCharacterisation}
\end{figure}

At very low field, electrons will simply pass through the THGEM holes and collect on the top plane, with no S2 light emitted. As the field increases, the electrons are now accelerated enough to excite the gaseous argon within the THGEM holes, resulting in the emission of S2 light. In the ``linear'' or ``pure electroluminescence'' regime, the number of emitted S2 photons is linearly proportional to the field applied across the THGEM. With further increase in the THGEM field, electrons gain enough energy to ionise the argon atoms, resulting in electron multiplication. In this ``exponential'' or ``electron multiplying'' regime, the number of emitted S2 photons exponentially increases with increasing THGEM field. 
\\
\\
Figure~\ref{fig:THGEMCharacterisation} shows the mean TPX3Cam ToT rate (calculated as the summed ToT of all hits in a run divided by the total duration of the run, and measured in ADU per second) as a function of the electric field across the THGEM. A single function - comprising a combination of linear and exponential functions - has been fitted to the data.

\noindent The linear and exponential regimes are clearly seen at low and high THGEM field respectively, with the cross-over point occurring at $\approx$ 24~kV/cm. There is a factor of $\approx$ 10 increase in the THGEM's S2 light production across the range of the exponential regime. It can also be observed that the TPX3Cam still has sensitivity deep into the linear regime, where the S2 light is minimal, and a subsequent benefit of this is that the THGEM can be operated at a very stable electric field during the entire duration of a run, with a significantly reduced possibility of discharges.

\subsection{Energy Calibration and Resolution}
\label{subsec:EnergyCalib}
Once the electron lifetime correction is applied, an energy calibration can be established. This is simply the conversion between the incident light intensity in ADU and the corresponding physical energy in MeV. As with the lifetime correction, through-going muons are ideal for calculating this calibration, since they are minimum-ionising particles (``MIPs'') with a well-known mean energy deposition rate, $\frac{dE}{dX}$, of 2.12~MeV/cm \cite{mipCalibration}.

\begin{figure}[ht]
\centering
\includegraphics[width=0.86\textwidth]{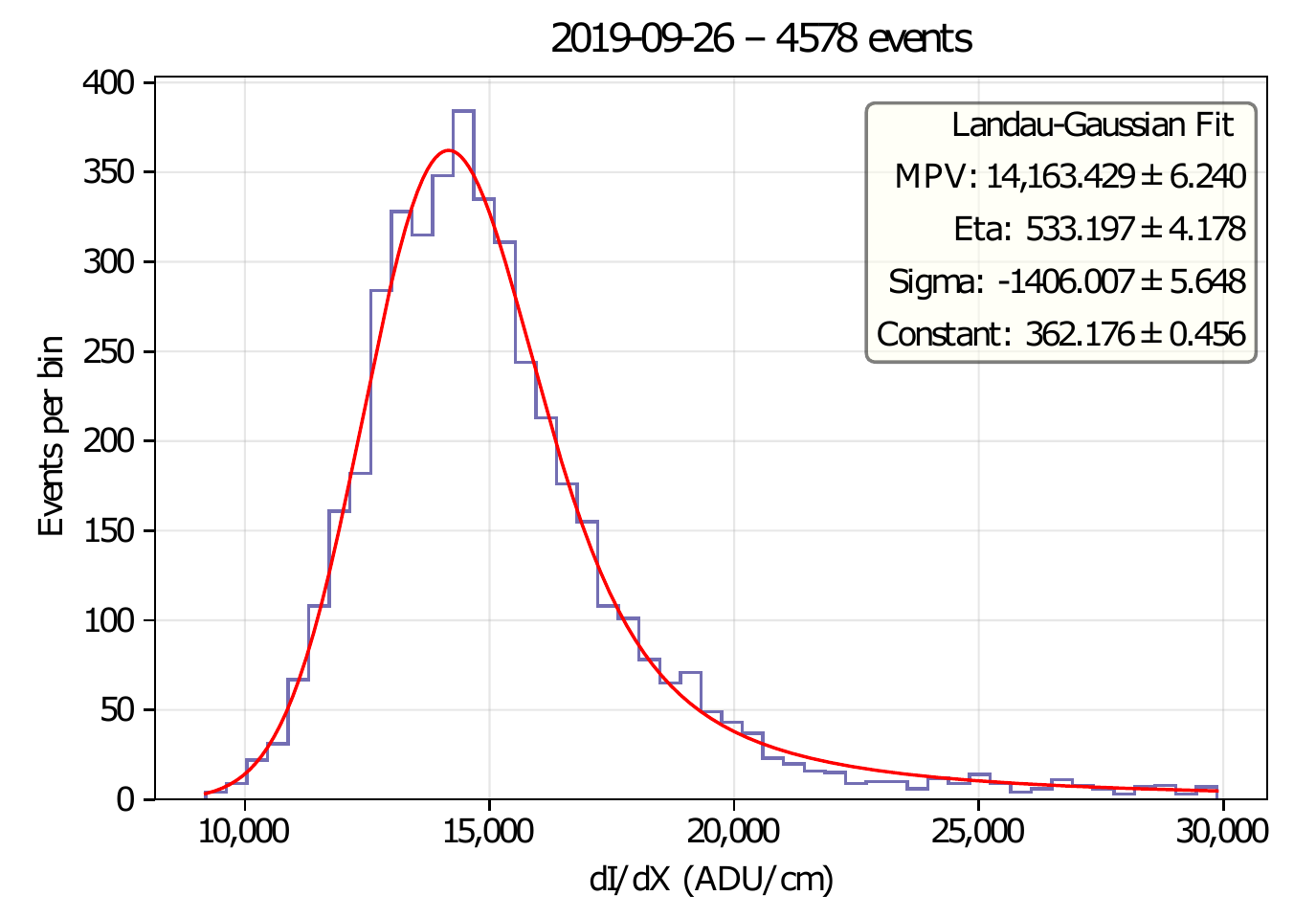}
\caption{The distribution of $\frac{dI}{dX}$ - the lifetime-corrected ToT summation - for a population of through-going muon tracks. A Landau-Gaussian convolution function (red) has been fitted to the entire distribution.}
\label{fig:energyCalibration}
\end{figure}

To measure the energy deposition of through-going muons, an event selection is performed on MIP-like events (which are well described by a single linear fit). The summed ToT is calculated across all hits which comprise each event, and this summation is divided by the 3D track length of the through-going track. This ratio is therefore equivalent to $\frac{dI}{dX}$, with units of ADU/cm. Figure~\ref{fig:energyCalibration} shows the distribution of this $\frac{dI}{dX}$ across a population of through-going muon tracks.\\

\noindent The calibration between ADU and MeV is found by equating the previously noted MIP energy deposition rate with the MPV of a Landau-Gaussian convolution function fitted to the $\frac{dI}{dX}$ distribution. For the dataset depicted in Figure~\ref{fig:energyCalibration}, which was taken at the nominal THGEM field of 29.45~kV/cm, the resulting calibration is $6681 \pm 3$ (stat.)~ADU per MeV. Similar calibrations have been found for other TPX3Cam datasets.

It is known that the response of the TPX3 ASIC is linear across the majority of its 1024~ADU ToT range, with a documented non-linearity at very low ToT values. Procedures for measuring and correcting this non-linearity have been documented by other groups \cite{ToTLinearity}, but have not been applied in this work. Further studies are planned to investigate the response of the TPX3Cam across a range of energy depositions, using other calibration sources in addition to cosmic muons.

The energy resolution, defined as the Landau (eta) and Gaussian (sigma) widths combined in quadrature and expressed as a fraction of the MPV, is approximately 11~\% for the dataset shown in Figure~\ref{fig:energyCalibration}. Improvements in energy resolution can be expected with increased electron lifetime, calibration of the TPX3 ASIC (i.e. corrections for non-linearities and pixel-to-pixel variations), as well as consideration of optical effects, such as lens vignetting. 

\subsection{Calorimetry using Stopping Muons}

\begin{figure}[ht!]
\centering
\includegraphics[trim={1.6cm 0cm 1.6cm 0cm}, clip, width=\textwidth]{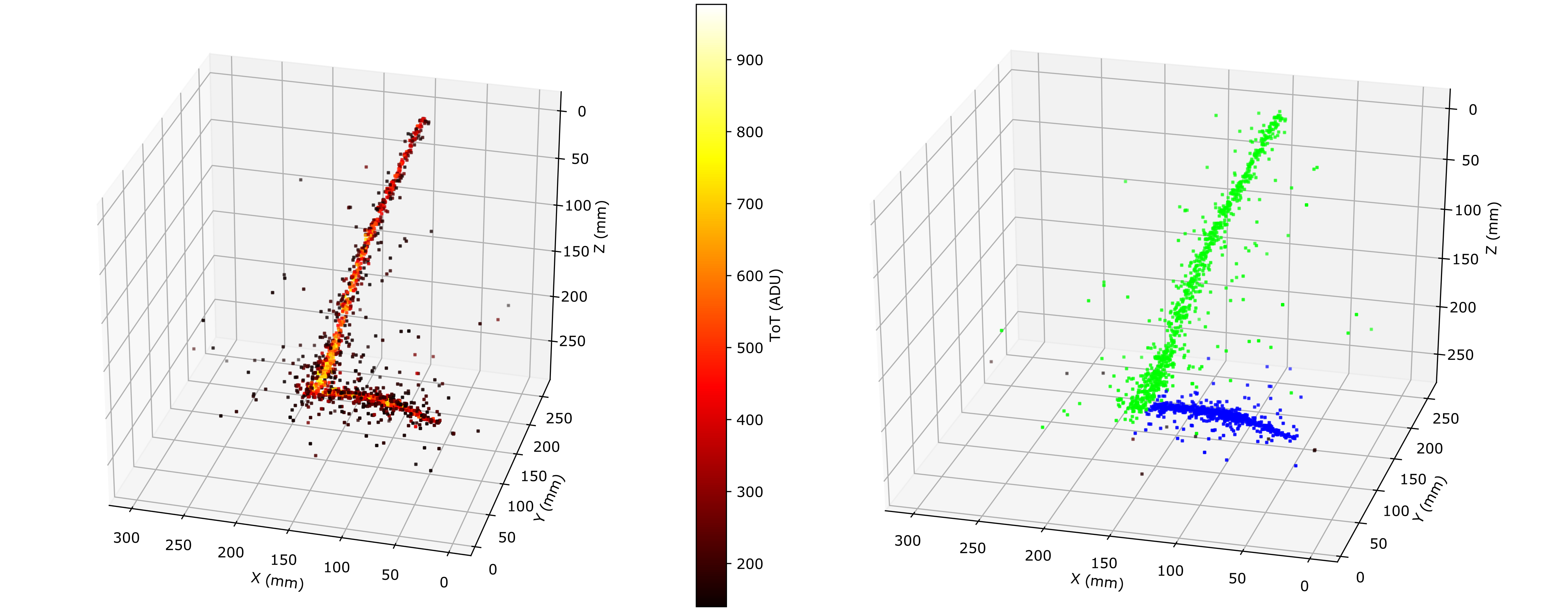}
\includegraphics[width=0.8\textwidth]{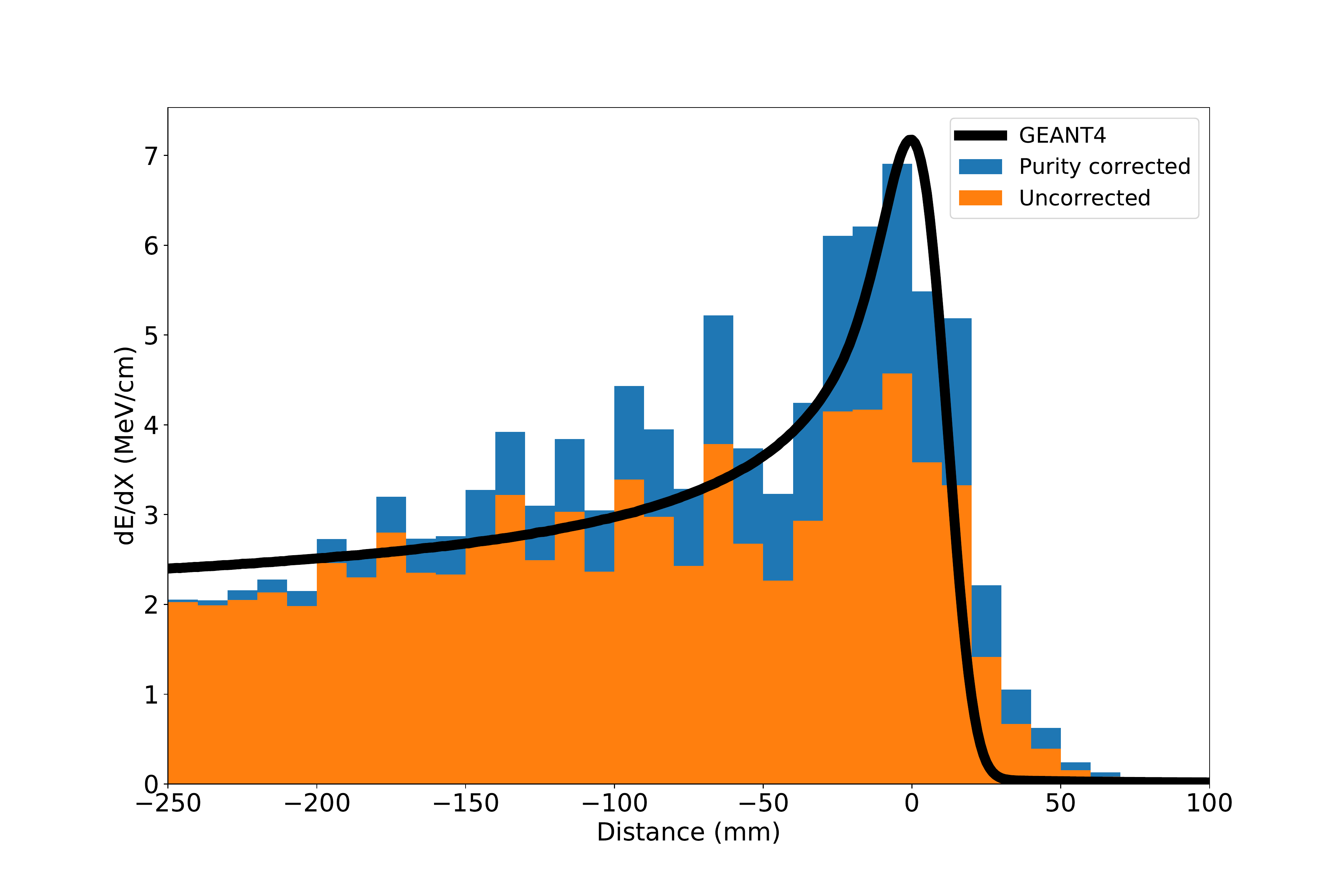}
\caption{Top left: a stopping muon track plotted with ToT color scale. Top right: the same stopping muon track plotted with hit assignment color scale. Hits assigned to the stopping muon are drawn in green, and hits assigned to the Michel electron are drawn in blue. Bottom: the muon track energy deposition as a function of distance. The zero distance of the track is assigned to the highest bin in the simulated Bragg peak. Negative distances indicate distance backwards along the stopping muon track.}
\label{fig:StoppingMuon1}
\end{figure}

\begin{figure}[ht!]
\centering
\includegraphics[trim={1.6cm 0cm 0.5cm 0cm}, clip, width=\textwidth]{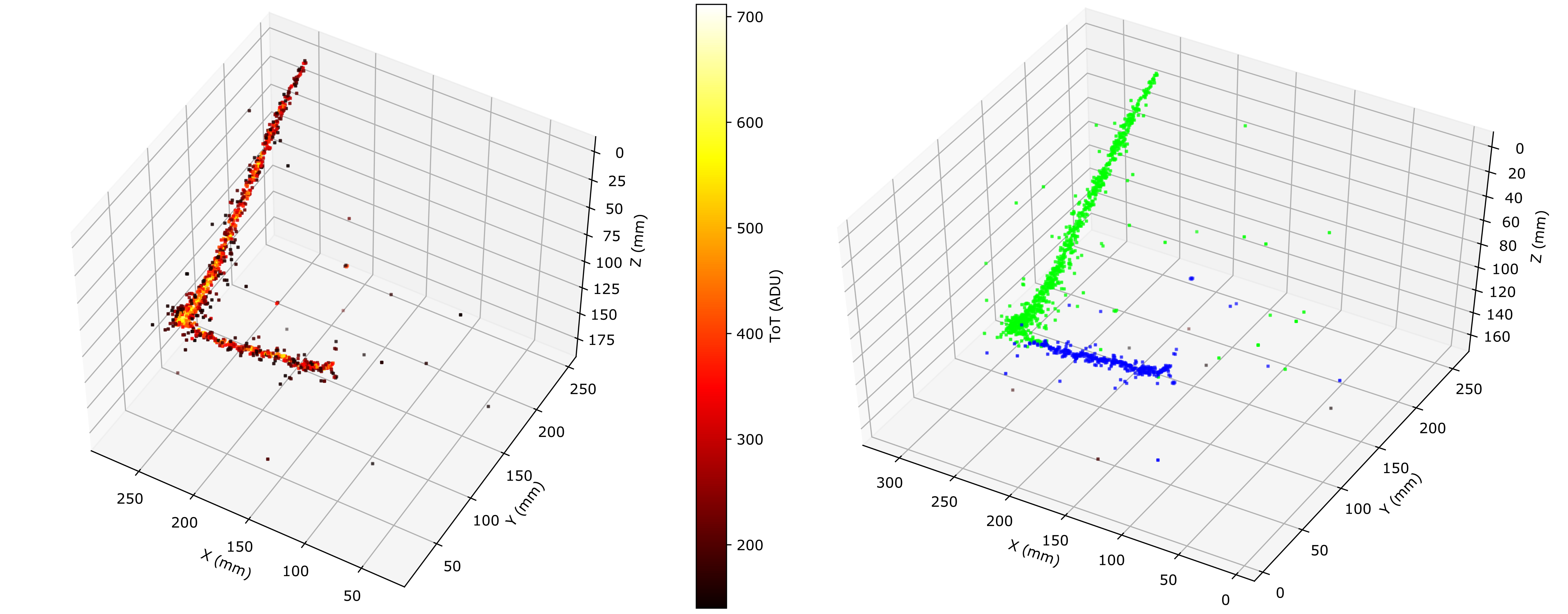}
\includegraphics[width=0.8\textwidth]{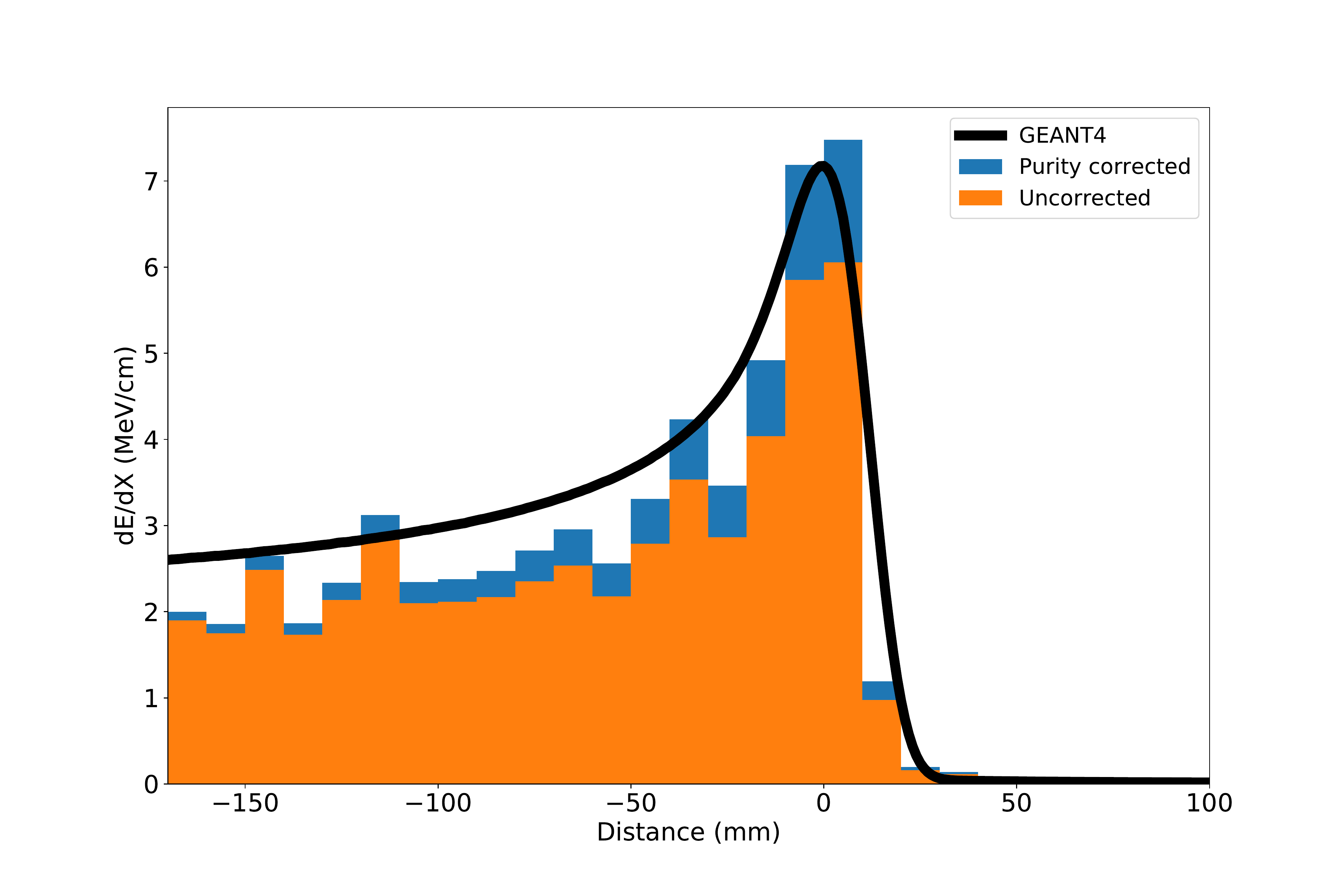}
\caption{Top left: a stopping muon track plotted with ToT color scale. Top right: the same stopping muon track plotted with hit assignment color scale. Hits assigned to the stopping muon are drawn in green, and hits assigned to the Michel electron are drawn in blue. Bottom: the muon track energy deposition as a function of distance. The zero distance of the track is assigned to the highest bin in the simulated Bragg peak. Negative distances indicate distance backwards along the stopping muon track.}
\label{fig:StoppingMuon2}
\end{figure}

For this study, a sample of two stopping muons was selected from the data.  These events were required to have entered the TPC through the THGEM.  This selection increases the possibility of identifying only those events with a relatively short drift distance, which in turn minimises the magnitude of purity correction required. For each event, a manual hit assignment was performed to separate those hits which comprise the muon track from those which are produced by the Michel electron. Once isolated, a straight line is fitted to the muon track. The energy profile of the stopping muon is reconstructed by summing the ToT of the muon hits along the fitted line in 1~cm wide bins. This profile (originally in units of ADU/cm) is then converted to MeV/cm using the energy calibrations described in Section~\ref{subsec:EnergyCalib}, allowing for direct comparison to simulation. 
\\
\\
The results of the analysis are shown in Figures \ref{fig:StoppingMuon1} and \ref{fig:StoppingMuon2}, with a GEANT4 model of the expected stopping muon energy profile overlaid on both. This model was produced by generating 100~MeV muons on the edge of a 1~$\times$~1~$\times$~1~$m^3$ cube of liquid argon. The cube is finely sliced in 1~mm segments perpendicular to the initial direction of the simulated muon. By recording the ionising energy deposition within each slice, the simulated energy profile of the stopping muon is generated. To minimise contamination from the Michel electron in the simulated stopping muon energy profile, only those ionising energy depositions within the first 5~ns of the event start time are recorded. This time has been found to be long enough for the muon to stop, but short enough that the muon is unlikely to decay, which would then produce the Michel electron track and its associated energy depositions. The developed GEANT4 model is in good agreement with models presented elsewhere \cite{mipCalibration, MicrobooneStoppingMuons}.

\section{Conclusions and Outlook}
The studies detailed in this paper have successfully demonstrated the imaging of cosmic muon-induced liquid argon interactions with a TPX3Cam mounted on the ARIADNE detector. The raw data from a TPX3Cam are natively 3D and zero suppressed, thus providing for straightforward event reconstruction as illustrated in the event gallery in Section~\ref{subsec:Gallery}. Taking advantage of the 1.6~ns time resolution of the camera, the drift velocity of the ionised electrons in liquid argon at the detector’s nominal 0.54~kV/cm electric field was determined to be 1.608~$\pm$~0.005~mm/$\mu$s (consistent with other experiments). After applying electron lifetime corrections to account for the argon purity, the energy calibration (allowing conversion of ADU to MeV) and resolution were determined using through-going muons. The energy resolution for the presented dataset was found to be approximately 11~\%. In addition, the energy deposition ($\frac{dE}{dX}$) as a function of distance for two stopping muons has been determined and compared to GEANT4 simulation, with good agreement demonstrated in both events. This represents a preliminary study, and further detector operation to accumulate higher statistics of stopping muons is expected to take place.
\\
\\
For this study, the TPX3Cam was mounted on a viewport that allowed for a view of just over a quarter of the THGEM active area. In light of this, one of the ongoing upgrades to the ARIADNE detector will be the placement of a TPX3Cam at the centre of the detector, with a custom made re-entrant viewport and a shorter focal length objective lens allowing for imaging of the entire active region of the THGEM. Additionally, use of a VUV-sensitive image intensifier will be investigated, which would eliminate the need for TPB wavelength shifter.

Preparations are also being made to instrument a larger volume TPC at the CERN Neutrino Platform using four TPX3Cam devices, viewing a total active area of 2~$\times$~2~m \cite{LOI}. This will allow for further characterisation of the system, as well as early indication and resolution of engineering challenges at larger scales.
\\
\\
The results discussed here, and the projected improvements from future optimisations, show great promise in terms of performance for kiloton-scale dual-phase LAr detectors, such as those planned for the DUNE programme. Furthermore, given that one camera can cover a large area (a single TPX3Cam is able to image a 1~$\times$~1~m area at 4~mm/pixel resolution), this system becomes very cost effective when considering these large-scale detectors. Moreover, in the imminent future the next generation TPX4 ASIC is expected to be released, which will have improved ToA (timing) and ToT (energy) resolution compared to TPX3, as well as a larger (448~$\times$~512) pixel array. Considering the 720~m$^{2}$ area of each of the proposed DUNE modules, an entire detector of this size could be read out with just 320 TPX4-based cameras, each looking at a field of view of 1.5~$\times$~1.5~m with $\approx$ 3~mm/pixel resolution. 

Taking into consideration the scientific, operational and cost benefits that this technology brings to the table, TPX-based optical readout of a dual phase LArTPC represents a serious option for the DUNE module of opportunity, for which innovative solutions are currently being sought. \\

\authorcontributions{Conceptualization, K.Mv., A.R., C.T. and J.V.; Funding acquisition, K.Mv. and C.T.; Investigation, A.L., K.Mj., K.Mv., B.P., A.R. and J.V.; Writing, A.L., K.Mj., K.Mv., B.P., A.R., C.T. and J.V. All authors have read and agreed to the published version of the manuscript.}

\funding{The ARIADNE program is funded by the European Research Council Grant No. 677927.}

\acknowledgments{The authors would like to thank the members of the Mechanical Workshop of the University of Liverpool's Physics Department, for their contributions and invaluable expertise.}

\conflictsofinterest{The authors declare no conflict of interest. The funders had no role in the design of the study; in the collection, analyses, or interpretation of data; in the writing of the manuscript, or in the decision to publish the results.}

\reftitle{References}

\end{document}